\documentclass{optica-article}
\journal{opticajournal} 
\articletype{Research Article}
\usepackage{lineno}

\usepackage{orcidlink} 

\begin{document}

\title{Variable coherence model for free-electron laser pulses}

\author{Austin Bartunek \orcidlink{0009-0004-0053-1837},\authormark{1,*} Nils H. Sommerfeld,\authormark{1,2} and Fran\c{c}ois Mauger \orcidlink{0000-0001-7555-6001}\authormark{1}}

\address{\authormark{1}Department of Physics and Astronomy, Louisiana State University, Baton Rouge, Louisiana 70803, USA\\
\authormark{2}Department of Physics, the Ohio State University, Columbus, Ohio 43210, USA}

\email{\authormark{*}abartu2@lsu.edu} 


\begin{abstract*} 
    We introduce the variable coherence model (VCM) for simulating free-electron laser (FEL) pulses generated through self-amplified spontaneous emission. Building on the established partial coherence model of [T. Pfeifer \textit{et. al}, Opt. Lett. 35, 3441 (2010)], we demonstrate that the implementation of a variable coherence width allows for continuous control over the pulses' characteristic noise, while keeping the average pulse parameters such as the bandwidth fixed. We demonstrate this through systematic statistical analyses of the intensity and number of sub-pulses in VCM pulses, in both time and frequency. In particular, we analyze how the sub-pulse statistics are affected by the coherence width parameter. We perform our analyses across three distinct regimes of FEL parameters and demonstrate how the VCM can generate pulses that range from maximally random to fully coherent. Finally, we illustrate the effect of the VCM variable coherence width on an absorption simulation.
\end{abstract*}

\section{Introduction} \label{intro}
In recent years, free electron lasers (FELs) have enabled the generation of light pulses with intensities previously inaccessible through tabletop sources alone~\cite{Duris2020,Ullrich2012,tera}. In particular, self-amplified spontaneous emission (SASE)~\cite{RevModPhys.88.015006} FELs can generate high-intensity sub-femtosecond soft x-ray pulses, thus enabling attosecond-science experiments targeting core-level electrons in atoms and molecules~\cite{core}. Recent FEL developments relevant for ultrafast science include pulse-pair generation that enable attosecond x-ray pump attosecond x-ray probe campaigns~\cite{Duris2020, Guo2024}, including two-color setups where each pulse has a different central photon energy~\cite{Cryan2022, Berrah2025}. Ongoing efforts also aim at pulse shaping~\cite{Li2024, Robles2025} in the form of chirped x-ray pulses.

\begin{figure}[htbp]
    \centering\includegraphics[width=.75\linewidth]{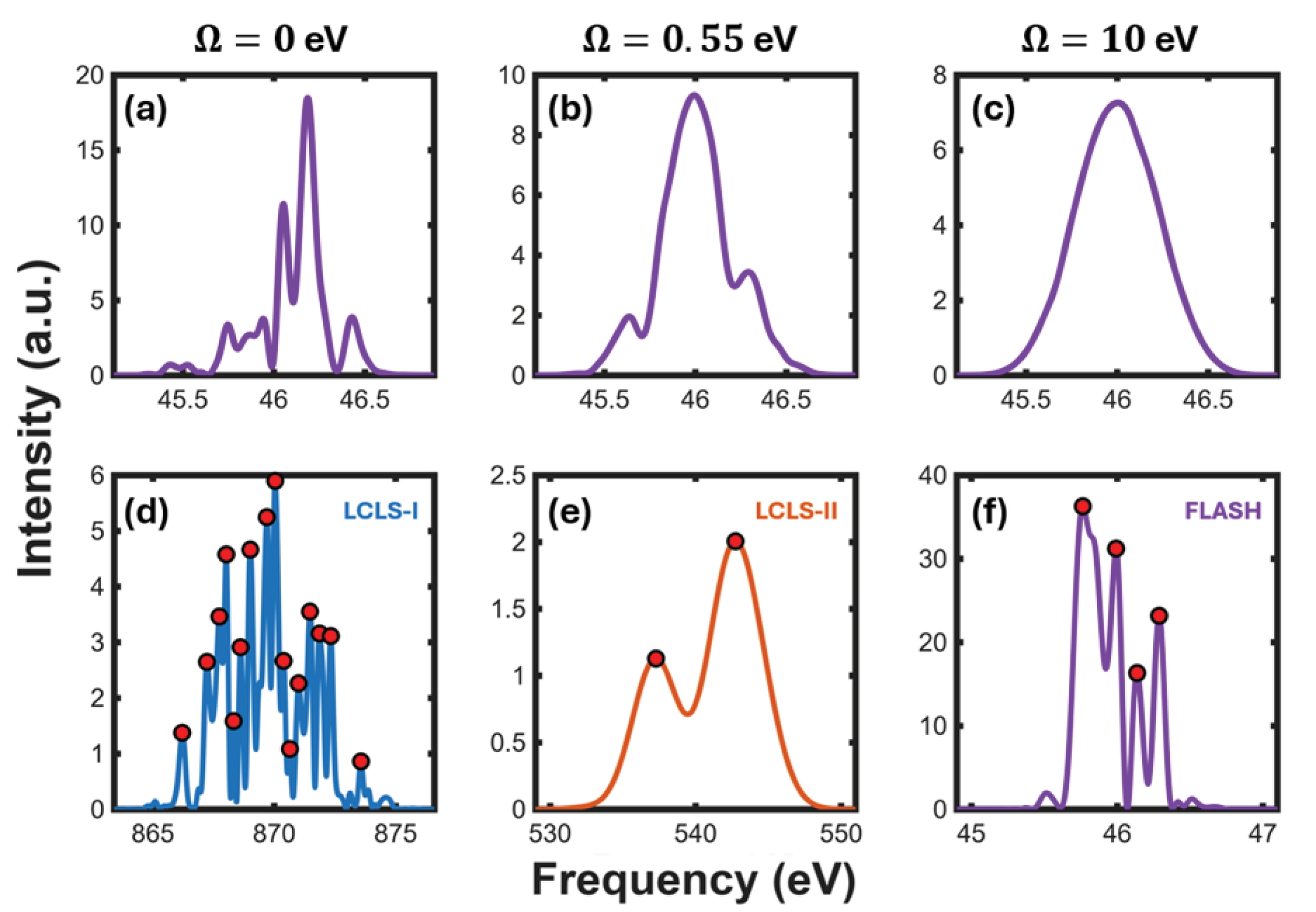}
    \caption{Examples of VCM pulses of equation~\eqref{eq:VCM_model} simulated using (a-c) three different coherence widths $\Omega$ and the FLASH parameters set, and (d-f) zero coherence widths for each parameter set in table~\ref{tab:FEL_table}.}
    \label{fig:ExamplePulses}
\end{figure}

A defining feature of SASE pulses is their stochasticity both within and between pulses, with typical pulses being characterized as a sequence of random intensity spikes (sub-pulses) with limited spectral and temporal coherence~\cite{BARLETTA201069,incoherence,SASE_stats}, as illustrated in figure~\ref{fig:ExamplePulses}~(a). Despite their shot-to-shot uniqueness, SASE pulses have proven to be an effective means to induce and probe a variety of processes in atoms and molecules~\cite{Berrah10072010, Sorokin_2006, PhysRevA.76.033416, PhysRevA.111.012808,Lozovoy:15}. To mitigate the impact of limited coherence and shot-to-shot jitter, a number of analysis methods have been developed such as covariance analysis within photoelectron and absorption spectroscopy~\cite{doi:10.1073/pnas.1821048116,Li_2021,correlation}, and ghost-imaging techniques within pump-probe schemes~\cite{PhysRevX.9.011045,XUV_GI}.

In this paper, we revisit the partial coherence model of Ref.~\cite{Pfeifer:10} to simulate SASE pulses with a controllable coherence width $\Omega$. Our variable coherence model (VCM) enables the simulation of pulses that range from minimally coherent, like those within the partial coherence model, to maximally coherent akin to Fourier-limited pulses. The coherence width dictates the magnitude of the coherence across a pulse's bandwidth, thereby allowing us to control its overall randomness while keeping the other parameters (central frequency and bandwidth) constant. Figure~\ref{fig:ExamplePulses} (a-c) illustrates the effect of introducing a variable coherence width, where we compare the spectral intensity profiles of three VCM pulses that only differ by their coherence widths. Clearly,  increasing the coherence width results in pulses with fewer sub-pulses, leading to a single Gaussian shape when the coherence width is very large. The VCM also provides an avenue for including pulse shaping in the model while retaining controllable stochasticity within and between pulses.

We quantify the VCM properties with systematic statistical analyses and comparisons of the intensity and number of sub-pulses contained within pulses of varying coherence width. We also investigate the correlation between sub-pulse statistics in the time and frequency domains. We apply our VCM statistical analyses to three distinct FEL parameter sets loosely matching the  LCLS-I, LCLS-II, and FLASH capabilities~\cite{Pfeifer:10,LCLS1_params1,LCLS1_params2,LCLS2_params} -- see table~\ref{tab:FEL_table} for the parameter values. Figure~\ref{fig:ExamplePulses} (d-f) shows an example of minimally coherent VCM pulses simulated for each parameter set. Finally, we illustrate the effect of a varying coherence width on the reshaping of an absorption feature in a model atomic system.

\begin{table}[htbp]
    \centering
    \begin{tabular}{||c|c|c|c||}
    \hline
         \textbf{FEL} &  $\boldsymbol{\omega_0} \textbf{ (eV)}$ &  $\boldsymbol{F}$ \textbf{(eV)} & $\boldsymbol{T}$ \textbf{(fs)} \\
         \hline\hline
         LCLS-I (long soft X-ray) & 870 & 4.4 & 10 \\
         \hline
         LCLS-II (short soft X-ray) & 540 & 5.5 & 0.5 \\
         \hline
         FLASH (XUV) & 46 & 0.55 & 25 \\
         \hline
    \end{tabular}
    \caption{Parameter sets for which we apply our VCM analyses, loosely corresponding to available modes of operation at LCLS-I (long soft X-ray),  LCLS-II (short soft X-ray), and FLASH (XUV). Each parameter set is defined by a central frequency $\omega_0$, average bandwidth $F$, and average pulse duration $T$ (FWHM in intensity).}
    \label{tab:FEL_table}
\end{table}

The structure of this paper is as follows: Section~\ref{methods} introduces the methods we use for our analyses, including the VCM in subsection~\ref{vcm} and the algorithm we use for sub-pulse detection in~\ref{sec:Methods:Sub-pulse_detection}. In section~\ref{statistics} we discuss the results of our statistical analyses, with sub-pulse analyses in the time and frequency domains in subsection~\ref{sec:Statistics:sub-pulse_analysis} as well as cross-correlation analyses of the time and frequency statistics in~\ref{sec:Statistics:cross-correlation}. Section~\ref{absorption} illustrates the effect of a varying coherence width on the reshaping of an absorption feature in a
model atomic system. Finally, we conclude our findings in section \ref{conclusion}.

\section{Methods} \label{methods}

In this section, we introduce the variable coherence model (VCM) and discuss its implementation for numerical simulations. Additionally, we establish the algorithms and criterion used for our sub-pulse analyses discussed in section~\ref{statistics}.

\subsection{Variable coherence model} \label{vcm}
Similar to the partial coherence model of Ref.~\cite{Pfeifer:10}, the VCM phenomenologically generates SASE pulses by randomly assigning the phases of the frequency components within the bandwidth of the pulse. Specifically, the electric field of a SASE pulse is given as 

\begin{equation} \label{eq:VCM_model}
    E(t)=NW_T(t-t_0)\mathcal{F}^{-1}\left[W_F(\omega-\omega_0)e^{i\phi(\omega)}\right](t),
\end{equation}
where $N$ is an overall scaling factor, 
$W_T$ (resp. $W_F$) is a window function that defines the average pulse envelope in time (resp. frequency), and $t_0$ (resp. $\omega_0$) is the central time (resp. frequency) of the pulse. $\mathcal{F}^{-1}$ denotes the inverse Fourier transform. We decompose the phase $\phi(\omega)$ into two components as
\begin{equation} \label{phases}
    \phi(\omega)=\phi_c(\omega)+\phi_v(\omega)
\end{equation}
where $\phi_c(\omega)=-\omega t_0$ is coherent part of the phase, ensuring that the pulse converges towards the Fourier limit at infinite coherence width ($\Omega\to\infty$). $\phi_v$ contains the stochastic part of the pulse and is responsible for the characteristic noise of SASE pulses. As pulse shaping capabilities like chirping become available in SASE FELs~\cite{pulse_shaping,Li2024,Robles2025}, one may incorporate such features in the coherent part of the phase $\phi_c$.

When the phases $\phi_v$ in equation~\eqref{phases} are sampled from a uniform random distribution over $[0,2\pi]$ we recover the partial coherence model~\cite{Pfeifer:10}, which creates a minimally coherent pulse. Throughout this paper, we refer to this as the zero coherence width limit ($\Omega=0$) of the VCM. Figure~\ref{fig:ExamplePulses}~(a,f) show examples of zero coherence width pulses for the ``FLASH'' set of parameters in table~\ref{tab:FEL_table}. We note that $\Omega=0$ still imposes some degree of coherence in the phases of neighboring frequencies, due to the convolution of the independent randomly selected $\phi(\omega)$ with the Fourier transform of the time window: $E(\omega)\propto\mathcal{F}[W_T]*(W_F e^{i\phi})$. Alternatively, a non-zero coherence width expands this local phase coherence across selectively larger portions of the spectrum. To do so, the VCM restricts the phase variation across frequencies by sampling $\phi_v$ from a L\'{e}vy process. Specifically, for a given frequency sampling $\omega_1,\omega_1,\ldots\omega_n$ we pick consecutive phases as
\begin{equation} \label{eq:random_phase_sampling}
    \phi_v(\omega_{k+1}) = \phi_v(\omega_k) + \pi \sqrt{\frac{\omega_{k+1}-\omega_k}{\Omega}} \Phi,
\end{equation}
where $\Phi$ is a random variable with zero mean value and unity standard deviation. The scaling factor $\sqrt{\omega_{k+1}-\omega_k}$ ensures that the stochastic spread of the phases between different VCM pulses is independent of the frequency sampling. For a random walk $\Phi\sim\sqrt{12}\ \mathcal{U}(-0.5,0.5)$, with $\mathcal{U}$ the uniform random distribution, and for Brownian motion $\Phi\sim\mathcal{N}(0,1)$, with $\mathcal{N}$ the normal distribution. Finally, to ensure the randomization of the overall pulses' carrier-envelope phase, we also add a global random phase sampled from $\mathcal{U}(0,2\pi)$ to all the $\phi_v$.

The coherence width $\Omega$ in equation~\eqref{eq:random_phase_sampling} quantifies the stochasticity of the phase contribution from $\phi_v$, with smaller $\Omega$ (larger stochasticity) leading to more randomness within and between VCM pulses. We have experimented with both the random walk and Brownian motion and found very similar results between them. All the results we show in this paper are obtained with the Brownian motion case. In the limit of infinite coherence width, $\phi_v$ becomes a constant across all frequencies, and the resulting pulse is fully coherent. Figure~\ref{fig:ExamplePulses}~(b) shows an example of a VCM pulse with a nonzero coherence width. Compared to the zero coherence width case in panel~(a), we see a clear reduction in the number of sub-pulses, reflecting the increased coherence between the pulse's frequency components. In the limit of very large coherence width, as in panel~(c), we observe a single broadband pulse with an envelope that closely matches the shape of the frequency window~$W_F$ in equation~\eqref{eq:VCM_model}. Small differences between $W_F$ and the observed pulse envelope come from the temporal window $W_T$ we impose on the overall pulse.

In our analyses, we set the normalization factor~$N$ in equation~\eqref{eq:VCM_model} such that all pulses have the same total energy $U$ given by
\begin{equation} \label{pulse_energy}
    U=\int\left|E(t)\right|^2dt=\int\left|E(\omega)\right|^2d\omega.
\end{equation}
Alternatively, shot-to-shot variations in the pulse energy can be incorporated by multiplying each normalized VCM pulse by a random variable sampled from the square root of the desired pulse-energy distribution. One may also easily apply alternative normalization conventions such as fixing the peak electric field amplitude or intensity.

\subsection{Sub-pulse detection} \label{sec:Methods:Sub-pulse_detection}

While VCM pulses have a well-defined and smooth structure on average, the random phases $\phi_v$ in equation~\eqref{phases} yield individual pulses comprised of a random number of local peaks with drastically varying intensities; we refer to these peaks as sub-pulses. Figure~\ref{fig:ExamplePulses} (d-f) shows three examples of typical zero coherence width pulses generated across the three regimes of interest from table \ref{tab:FEL_table}. For each pulse, the peaks of the constituent sub-pulses are identified with a red marker.  In the time domain, VCM pulses are comprised of fast oscillations at the central frequency~$\omega_0$ modulated by a slowly-varying envelope. In our sub-pulse analyses, we ignore these fast oscillations and only consider the slowly-varying envelopes. From figure~\ref{fig:ExamplePulses}, we identify that the number and intensity of sub-pulses are extremely sensitive to the pulse parameters used to simulate them. In section~\ref{statistics}, we investigate the statistical properties of the sub-pulses produced by the VCM, specifically regarding their dependence on the coherence width both in the time and frequency domains for each of the parameter sets in table~\ref{tab:FEL_table}.

To study the statistics associated with the sub-pulses, we first require a means for automatically capturing their locations and intensities within individual VCM pulses. The peak of each sub-pulse can be identified as a local maximum within the full pulse, which we obtain using the \texttt{findpeaks} function in \texttt{MATLAB}'s signal processing toolbox~\cite{MATLAB_findpeaks}. We note that the temporal and spectral windows $W_T$ and $W_F$ in equation~\eqref{eq:VCM_model} have the effect of drastically suppressing the intensities of sub-pulses that lie near the tails of any given pulse, so we discard these weaker sub-pulses from our statistical analyses by setting a threshold on the minimum peak heights. The stochastic nature of the pulse generation means that some sub-pulses may feature small local maxima within their temporal/spectral amplitude without these maxima being associated with a different sub-pulse. We similarly discard these local maxima by thresholding their prominence compared to neighboring peaks. Specifically, for all parameter sets listed in table~\ref{tab:FEL_table} and both in the time and frequency domains,  we disregard any sub-pulses that are less than 10\% of the intensity of the most intense sub-pulse occurring within the same VCM pulse. Then, for all parameter sets, we demand that the peak prominence is at least 4.5\% of the maximum sub-pulse intensity in the frequency domain, and 10\% in the time domain. The larger prominence threshold in the time domain is chosen due to an increased amount of small local maxima within the sub-pulses' envelopes. Examples of qualifying sub-pulses are highlighted with red markers in figure~\ref{fig:ExamplePulses}. We have checked that our results are robust with small changes in the threshold criteria we use in our analyses.

For each parameter set listed in table \ref{tab:FEL_table} and for each coherence width, we perform our sub-pulse analyses over 10,000 VCM pulses for which we find that all our statistics and distributions have converged. Additionally, we have made the arbitrary choice to fix the total energy of every VCM pulse at $U=1$ a.u. A different energy value would systematically scale the amplitude of all the VCM pulses, leading to a scaling of the sub-pulse intensities without otherwise affecting the shapes of the distributions we show in the following section.

\section{Statistical Analyses}\label{statistics}

\subsection{Sub-pulse statistics} \label{sec:Statistics:sub-pulse_analysis}
\begin{figure}[htbp]
    \centering
    \includegraphics[width=.75\linewidth]{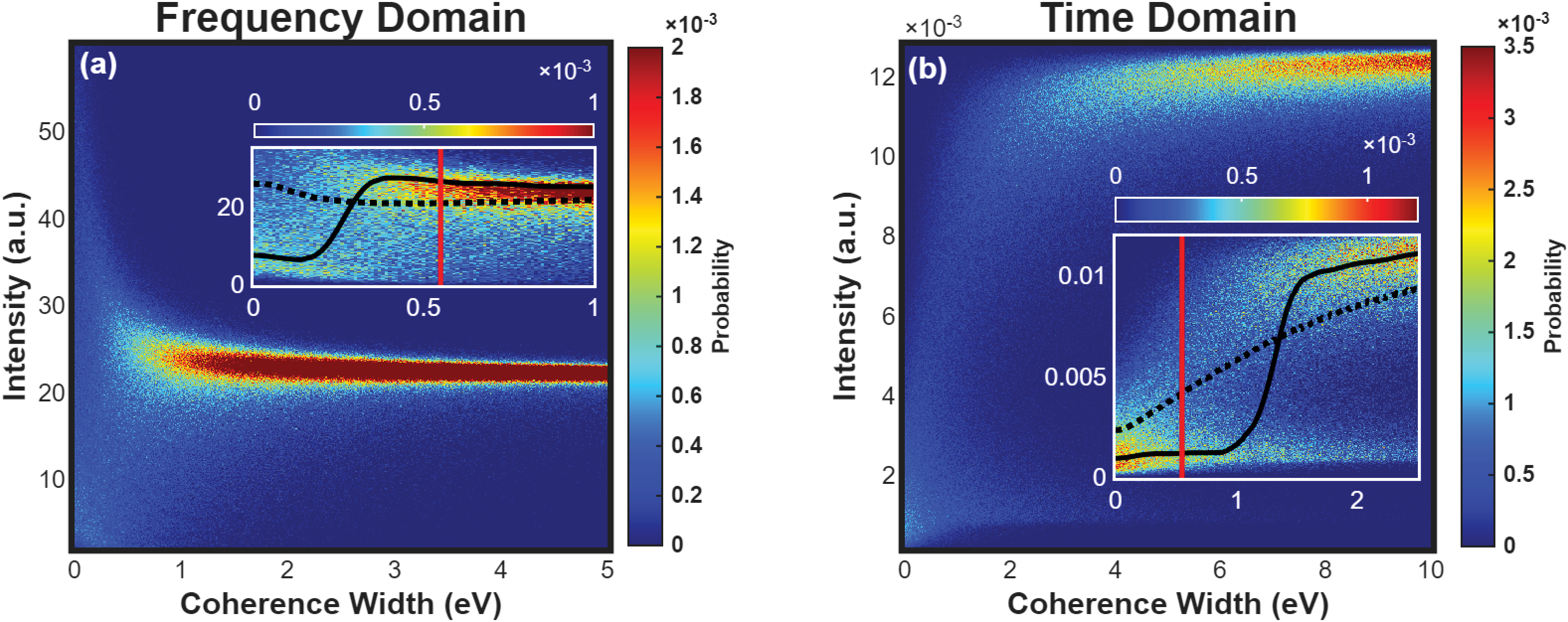}
    \caption{Simulated sub-pulse intensity probability distributions as a function of the coherence width for the FLASH parameters of table~\ref{tab:FEL_table} in the (a) frequency and (b) time domains. In each panel, the inset zooms in on the lower coherence widths where the probability transitions from low to high average value distributions. The dotted and solid curves mark the mean and mode of each distribution at a fixed coherence width, respectively. The vertical red line marks the bandwidth of the FLASH parameter set (0.55 eV).}
    \label{fig:IntColor}
\end{figure}

To begin our analyses, we examine the distributions of simulated sub-pulse intensities in the frequency and time domains and show the results for the FLASH parameters in figure~\ref{fig:IntColor}. Panel~(a) shows the probability of generating a sub-pulse with some particular intensity as a function of the coherence width and the inset depicts a scaled view across a smaller range of values. For a low coherence width, below $\approx$~0.25~eV or about half the pulses' bandwidth, we see a large and non-uniform spread of possible sub-pulse intensities highlighted by the significant difference between the mean (dotted curve in the inset) and mode (solid) of the sub-pulse distribution. As the coherence width further increases, we see a narrowing of the distribution that converges towards the fully-coherent Fourier-limited case. This behavior at large coherence widths is precisely what we expect from a fully coherent pulse: one prominent peak with a consistent maximum intensity. In panel~(b), we show the corresponding sub-pulse intensity distribution in the time domain. In this case, most of the sub-pulses at low coherence widths below $\approx$~0.25~eV are concentrated towards smaller intensities rather than having a large spread as they did in the frequency domain -- here the mean and mode (dotted and solid curves) of the sub-pulse distribution are close to each other. Then, as the coherence width is increased, we see that the most probable sub-pulse intensity gradually increases towards an asymptotic value just like it did in the frequency domain. The supplemental document presents similar sub-pulse intensity analyses simulated for the LCLS-I and LCLS-II sets of parameters. For LCLS-I, we see very similar behavior to the FLASH case, but the LCLS-II results are more distinct. Nonetheless, in all cases we observe a convergence towards asymptotic sub-pulse intensities. 

A notable difference between the time and frequency sub-pulse intensity distributions of figure~\ref{fig:IntColor} is how fast they converge towards the fully-coherent limit when the coherence width is increased. For a coherence width equal to the FEL bandwidth, we see in panel~(a) that the distribution in the frequency domain is predominantly converged around its final value, whereas for the time domain in~(b), the distribution is still biased towards lower intensities until the coherence width is $\approx$~1.5~eV, and not fully converged until the coherence width is near ten times the FEL bandwidth. Looking at individual VCM pulses, we attribute the difference between the time and frequency to how the pulses' coherence width affects their structures: in the frequency domain, increasing the coherence width results in the sub-pulses progressively moving closer to the peak of $W_F$ from equation~\ref{eq:VCM_model}, naturally increasing their intensities -- see for instance figure~\ref{fig:ExamplePulses} (a-c). In the time domain on the other hand, increasing the coherence width leads to a narrowing of the pulse around $t_0$, which means that most of the sub-pulses manifest as lower-intensity pedestals surrounding the narrow central peak. For intermediate coherence widths, these pedestals are very prevalent, and thus weigh down the most probable sub-pulse intensity, resulting in a slower transition towards the asymptotic intensity compared to the frequency domain.

\begin{figure}[htbp]
    \centering\includegraphics[width=.8\linewidth]{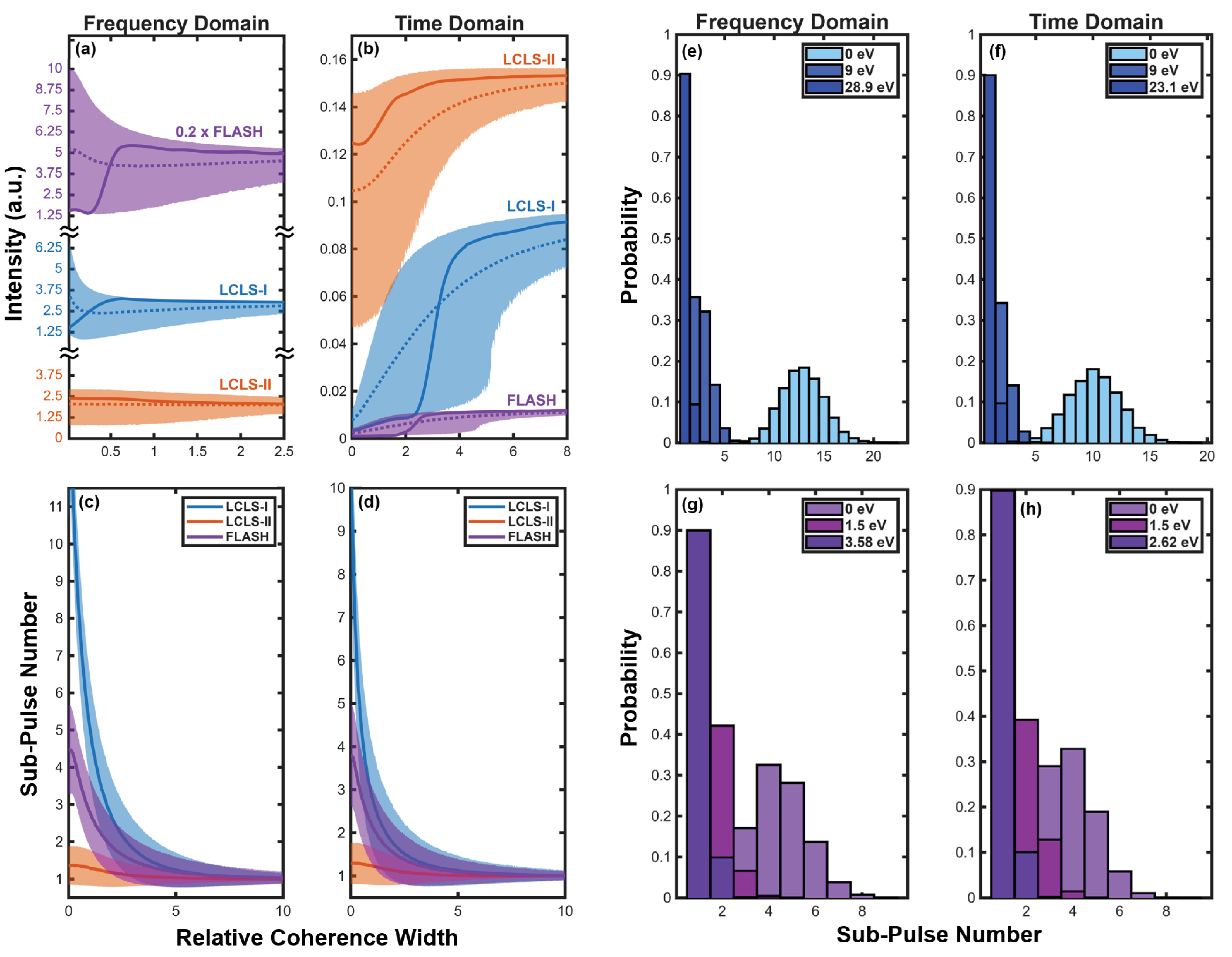}
    \caption{(a-b) Simulated mean sub-pulse intensities (dotted) and sub-pulse modes (solid) as functions of the relative coherence width, defined as the ratio between the coherence width and average pulse bandwidth, for the LCLS-I (blue), LCLS-II (orange), and FLASH (purple) FEL parameter sets of table~\ref{tab:FEL_table} in the (a) frequency and (b) time domains. For each data set, the upper and lower bounds of the shaded regions represent the 90$^{th}$ and 10$^{th}$ percentiles of the sub-pulse intensity distributions, respectively. In panel~(a), we scaled the FLASH results for visual clarity.
    (c,d) Simulated average number of sub-pulses (curves) plotted against the relative coherence width for the LCLS-I (blue), LCLS-II (orange), and FLASH (purple) FEL parameter sets of table~\ref{tab:FEL_table} in the (c) frequency and (d) time domains. For each curve, the shaded regions represent $\pm$1 standard deviation around mean value. (e-h) Histograms of the sub-pulse number distributions simulated for three exemplary values of the coherence width for the (e,f) LCLS-I (blue) and (g,h) FLASH (purple) parameter sets in the (e,g) frequency and (f,h) time domains.}
    \label{fig:IntensityNumberStats}
\end{figure}

We investigate the effects of the coherence width on the sub-pulse intensities simulated with the other FEL parameter sets of table~\ref{tab:FEL_table} in figure~\ref{fig:IntensityNumberStats} (a,b), where we show the average sub-pulse intensities and modes as a function of the coherence width for both the (a)~frequency and (b)~time domains -- also shown as the dotted and solid curves in the insets of figure \ref{fig:IntColor}, respectively. For all curves, the lower and upper bounds of the shaded regions represent the 10$^{th}$ and $90^{th}$ percentiles of the distributions. In the frequency domain of panel~(a), we recognize a similar behavior between the mean curves describing the FLASH and LCLS-I parameter sets while the LCLS-II case produces distinct results. For FLASH and LCLS-I, the average sub-pulse intensity first decreases when increasing the coherence width from zero. We attribute this feature to the large spread of possible intensities that initially weights the average towards a larger value at zero coherence width before clustering near lower intensities for small nonzero coherence widths. After reaching a shallow minimum, the average sub-pulse intensity converges towards its asymptotic Fourier-limited value, due to the concentration of sub-pulses towards the pulses' central frequency. On the other hand, the LCLS-II average sub-pulse intensity is maximized at zero coherence width, followed by a slow but consistent decline until it reaches its asymptotic value. In the time domain of panel~(b), we see that all parameter sets display similar qualitative behavior, with each average sub-pulse intensity curve monotonically increasing towards its asymptotic Fourier-limited value. Comparing the sub-pulse intensity distributions between the two panels (shaded areas), we see that broader distributions in the frequency domain are associated with narrower ones in the time domain, and vice versa, as a result of the uncertainty principle. We also see that the range of coherence widths required to reach the Fourier-limited asymptotic value is considerably larger in the time domain than in the frequency domain, as we observed in figure~\ref{fig:IntColor} for the FLASH case.

Next, we investigate the coherence width's effects on the number of sub-pulses contained within a given VCM pulse. Figure~\ref{fig:IntensityNumberStats} (c,d) compares how the average number of sub-pulses depends on the coherence width, simulated across all parameter sets and in both the (c) frequency and (d) time domains. In all cases, we see that the average number of sub-pulses is maximized at zero coherence width. Then, we find that increasing the coherence width results in the average number of sub-pulses monotonically decreasing towards one, just like we would expect from a fully-coherent pulse. In the FLASH and LCLS-I cases, we further find that the decrease follows an exponential law. Like for the sub-pulse intensities of figure~\ref{fig:IntensityNumberStats} (a,b), we see that the distributions of sub-pulse numbers narrow with increasing coherence width -- see the shaded areas at $\pm1$ standard deviation around the average.

Figure~\ref{fig:IntensityNumberStats}~(e-h) shows individual sub-pulse number distributions at three exemplary values of the coherence width simulated with the LCLS-I (top row) and FLASH (bottom row) parameter sets within both the frequency (left column) and time (right column) domains. At zero coherence width, we find that both parameter sets yield similar number distributions between the time and frequency domains, with a mean of $\approx11$ sub-pulses for LCLS-I and $\approx4$ sub-pulses for FLASH. As the coherence width increases well beyond the bandwidth of each parameter set, the distributions approach the single-sub-pulse limit for smaller coherence widths in the time domain than in frequency. For each panel~(e-h), the leftmost distributions correspond to the minimal coherence width required such that at least 90\% of VCM pulses contain a single sub-pulse, thereby reliably producing pulses that are effectively coherent. For the LCLS-I parameter set, this coherence width is 28.9~eV ($\approx5.2$ times the average pulse bandwidth) in the frequency domain and 23.1~eV ($\approx4.2$ times) in the time domain, whereas for FLASH it is  3.58~eV ($\approx6.5$ times) in the frequency domain and 2.62~eV ($\approx4.7$ times) in the time domain. Similar to the sub-pulse intensity, we attribute the differences in the minimum coherence widths required for convergence of the sub-pulse number in time vs frequency to the effect of increasing the coherence width on the shapes of the pulses in the time (narrow central peak with pedestals) and frequency (limited effect on the average pulse envelope) domains.

\subsection{Joint Probability Distributions} \label{sec:Statistics:cross-correlation}
For all of the results presented thus far, we have divided our discussion of the sub-pulse properties between the time and frequency domains. In doing so, we have treated the individual VCM pulses as being unconcerned with their counterparts in either domain, thereby ignoring any potential correlations between them. We now shift our focus to investigate the extent to which these correlations exist. 

\begin{figure}[htbp]
    \centering
    \includegraphics[width=\linewidth]{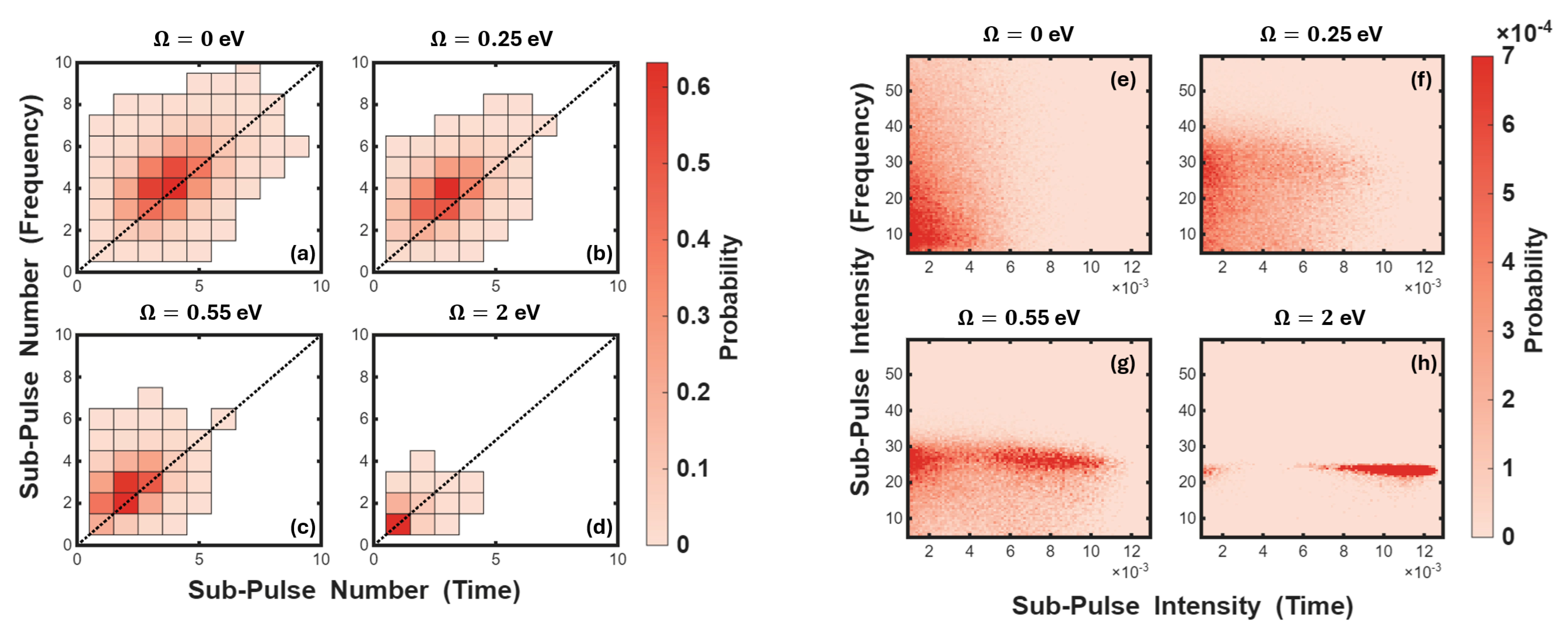}
    \caption{Simulated joint probably distributions for the sub-pulse (a-d) numbers and (e-h) intensities in the time and frequency domains for VCM pulses with increasing coherence widths -- see the titles. All panels are obtained with the FLASH set of parameters in table~\ref{tab:FEL_table}.}
    \label{fig:JointHistoFlash}
\end{figure}

Figure~\ref{fig:JointHistoFlash} (a-d) shows a series of joint probability distributions that describe the likelihood of simulating a FLASH pulse with a particular number of sub-pulses in both the time and frequency domains. In the zero coherence width case of panel~(a), we observe a weak, positive correlation between time and frequency, with the most likely number of sub-pulses being 4 in both domains -- consistent with our findings in figure~\ref{fig:IntensityNumberStats} (g,h). Importantly, the large spread in the joint sub-pulse-number probability distribution suggest that there is little correlation, if any, between the sub-pulses in the two domains. We attribute the slightly increased likelihood of generating more sub-pulses in the frequency domain than the time domain to our choices of a slightly higher prominence criterion in time in our sub-pulse detection -- see Sec.~\ref{sec:Methods:Sub-pulse_detection}. As the coherence width is increased towards the bandwidth in panels~(b-c), we see that the distribution begins to narrow along the main diagonal, indicating increasing correlation strength. Comparing the four panels, we also see that larger coherence widths result in fewer sub-pulses within both domains, which is again consistent with the results of figure~\ref{fig:IntensityNumberStats}. Put together, these results demonstrate that the number of sub-pulses in the time and frequency domains is mildly correlated for the FLASH parameters, particularly for small coherence widths. We observe qualitatively similar trends in our simulations with the LCLS-I and LCLS-II parameters, with analogous joint probability distributions to figure~\ref{fig:JointHistoFlash} shown in the supplemental document. For LCLS-II, the joint probability distribution is nearly symmetric and already concentrated towards mostly one or two sub-pulses in both the time and frequency domains, reflecting the reduced duration of pulses with respect to the average bandwidth.

The correlation analysis for the intensities of sub-pulses in the time versus frequency domain is less straightforward than for the sub-pulse number, as there is no obvious one-to-one mapping between sub-pulses in either domains -- recall that many VCM pulses have a different number of time vs frequency sub-pulses. Instead, we consider all possible pairings of time-frequency sub-pulse intensities within each VCM pulse and show the corresponding joint probability distribution in figure~\ref{fig:JointHistoFlash} (e-h). At zero coherence width in panel~(e), the bulk of intensity pairs lie in regions of low intensity for both domains without exhibiting substantial correlation with one another. As the coherence width is increased to 0.25~eV in panel~(f), the distribution begins to take on a more defined structure, with the bulk of intensity pairings still concentrated towards lower intensities in time but higher intensities in frequency. This observation is consistent with our understanding of the coherence width's effect on a pulse's structure: the time domain requires larger coherence widths for the sub-pulse intensities to reach their asymptotic value due to the appearance of lower-intensity pedestals that weigh down the average -- see the discussion around figure~\ref{fig:IntColor}. When the coherence width equals the average bandwidth in panel~(g), we see that most of the sub-pulse intensities in frequency are concentrated around the Fourier-limited limit while we still have a broad distribution of intensities in time. As the coherence width is further increased beyond the bandwidth in panel~(h), the sub-pulse intensities in time begin to approach their asymptotic Fourier-limited value. Similar to the sub-pulse number in panels~(a-d), we generally observe a weak correlation between sub-pulses in the time vs frequency domains. Instead, the joint probability distributions highlight the different rate at which sub-pulses in each domain approach their Fourier-limited value as the coherence width is increased.

The supplemental document presents the simulated joint probability distributions of sub-pulse intensities in the time versus frequency domains for the LCLS-I and LCLS-II parameters. For the LCLS-I parameter set, we observe qualitatively similar results to the FLASH case from figure~\ref{fig:JointHistoFlash}. On the other hand, the LCLS-II parameter set yields a very distinct behavior: irrespective of the coherence width, we observe a strong (anti-)correlation between the time and frequency sub-pulse intensities, which tend to cluster around their Fourier-limited asymptotic values. We attribute these results to the limited number of sub-pulses -- mostly one or two per pulse -- again, due to the reduced pulse duration compared to the average bandwidth.

\section{Absorption Calculations}\label{absorption}
In this section, we illustrate how the coherence width of VCM pulses affects nonlinear absorption spectrum properties of a model atomic system. In doing so, we bridge the absorption properties from minimally coherent pulses, akin to what is available in (x-ray)FEL campaigns, to fully coherent ones, as is often considered in theory.

\begin{figure}[htbp]
    \centering\includegraphics[width=\linewidth]{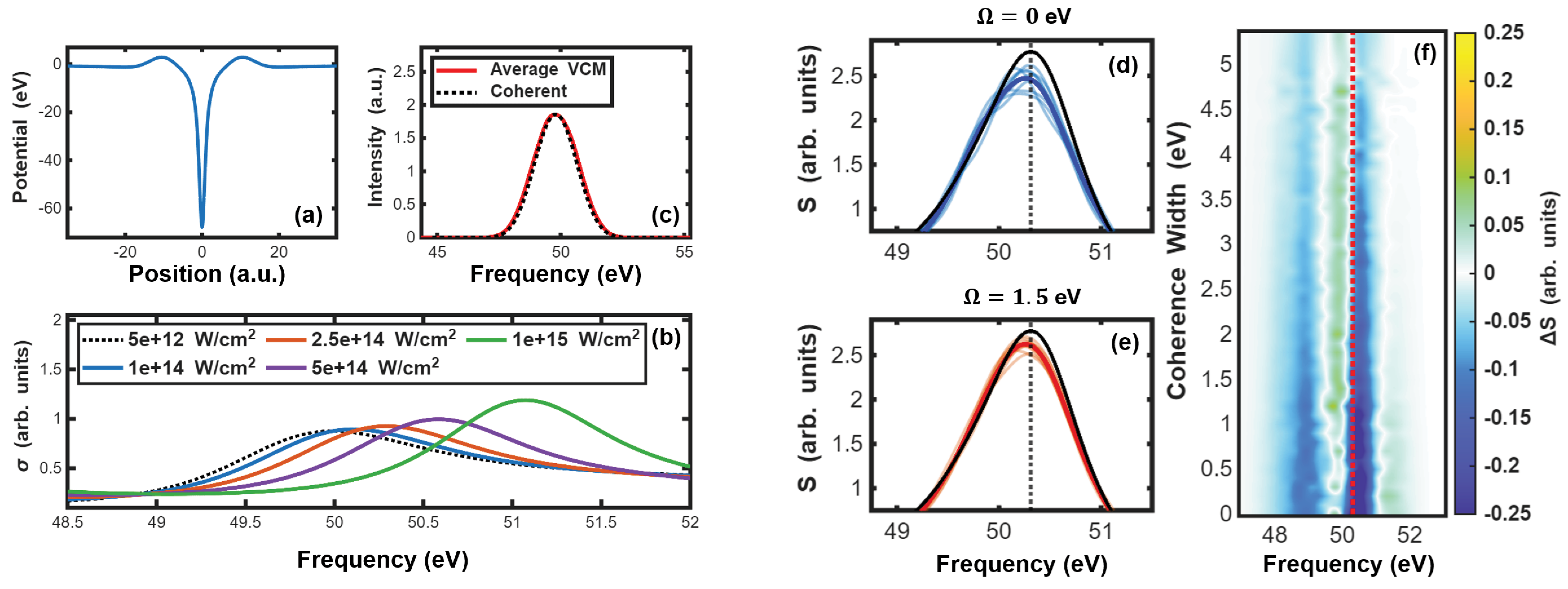} 
    \caption{(a) Atomic potential used in our absorption simulations. (b) Absorption cross-sections obtained from coherent  cos$^4$ pulses across a range of intensities. (c) Spectral intensity for the~$4.65\times10^{14}$~W/cm$^2$ peak intensity coherent cos$^4$ pulse (dotted curve) and matching average VCM pulse with zero coherence width (solid).
    (d,e) Average response functions of equation~\eqref{cross_section} for VCM pulses with the coherence width specified in the titles. In each panel, the opaque thick curve corresponds to the grand averages while transparent curves are batch averages over 250 VCM pulses only. For reference, the thin black curve corresponds to the average response function calculated with a coherence width of 10~eV -- very close to the fully coherent Fourier limit. (f) Difference between the grand average response functions and reference average at 10~eV as a function of the VCM coherence width. Horizontal sections at 0 and 1.5~eV correspond to the difference between the thick opaque and thin black curves in panels~(d,e). In all panels, the vertical dotted line at 50.31~eV marks the frequency at which the reference average response function is maximized.}
    \label{fig:Absorption}
\end{figure}

For our absorption simulations, we consider a one-dimensional atomic model with 4 active electrons using density-functional theory~(DFT). We use a soft-Coulomb potential $-Z/\sqrt{x^2+a^2}$~\cite{Javanainen1988}, with $Z=4$ and $a=0.945$~a.u., to which we add two Gaussians on either side of the potential to structure the absorption cross section close to the ionization threshold~\cite{shape_res}, akin to a shape resonance. Each Gaussian is located $10$~a.u. from the atomic center with a height of 0.2~eV and a width of 3.5~a.u. We show the resulting atomic potential in figure~\ref{fig:Absorption}~(a). We calculate the electronic ground state and perform the subsequent time-dependent DFT (TDDFT) simulations using the QMol-grid package~\cite{QMol} with an LDA Slater-exchange potential~\cite{LDA}.

Starting from the ground state, we calculate the response function~$S$ and absorption cross-section~$\sigma$ from the time-dependent dipole acceleration $\ddot{d}(t)$ as~\cite{BECK2015119,Wu2016}
\begin{equation} \label{cross_section}
     \sigma(\omega) = 
        -\frac{1}{\omega}\Im\left\{\frac{\mathcal{F}[\ddot{d}](\omega)}{\mathcal{F}[E](\omega)}\right\}
    \qquad \text{and} \qquad
    S(\omega)  = 
        -\frac{2}{\omega}\Im\left\{\mathcal{F}[\ddot{d}](\omega)\mathcal{F}[E](\omega)^*\right\},
\end{equation}
where $\Im$ denotes the imaginary part and $^*$ is the complex conjugate. Figure~\ref{fig:Absorption}~(b) shows the absorption cross-sections for our atomic model calculated with fully coherent cos$^4$ pulses with $\omega_0=49.8$~eV central frequency, 1~fs\ pulse duration ($\approx$~2~eV bandwidth) FWHM intensity, and various peak field intensities. For an intensity of $5\times10^{12}$~W/cm$^2$ (dotted curve), we obtain the linear response of the system where the absorption cross-section is independent of the intensity and features a local maximum at $49.8$~eV. As the intensity is increased above $10^{14}$ W/cm$^2$ (solid curves), we observe progressive reshaping of the absorption feature due to non-linear effects, with the local maximum progressively shifting to higher photon energies. Here, we are interested in the nonlinear regime and select a peak intensity of~$4.65\times10^{14}$~W/cm$^2$ moving forward. Specifically, in the VCM model of equation~\eqref{eq:VCM_model} with zero coherence width, we select the defining parameters of $W_T$ and $W_F$ such that the average VCM spectral intensity closely matches that of the fully coherent cos$^4$ reference -- see figure~\ref{fig:Absorption}~(c). The corresponding parameters are $T=5$ fs, $F=2$~eV, $\omega_0=49.8$~eV, and $U=0.3064$~a.u. We then vary the coherence width while keeping the other parameters fixed.

The presence of $\mathcal{F}[E]$ in the denominator of $\sigma$ in equation~\eqref{cross_section} makes the absorption cross-section from individual VCM pulses highly sensitive to low-intensity dips that are characteristic of small coherence widths. Instead, we consider the response function $S$ and show the results in figure~\ref{fig:Absorption} (d-f). Panel~(d) shows the results for minimally coherent pulses with zero coherence width. The thin transparent curves, obtained by averaging the response function over 250 VCM pulses, yield significant fluctuation between batches, reflecting the large shot-to-shot fluctuations in the pulses -- see the discussion in section~\ref{statistics}. After averaging over 4000~pulses, we obtain a smooth average response function that is notably different from the result at large coherence width -- compare the thick opaque and thin black curves. In particular, we see that the maximum in the response function at 50.31~eV is about 10\% smaller for the zero-coherence width curve compared to the 10~eV reference. Panel~(e) shows a similar analysis for a coherence width of 1.5~eV, slightly below the pulses' average bandwidth. Compared to the zero coherence width case, we see that the 250 VCM pulse batches yield average response functions that are closer to the grand average, and that the grand average, which is now converged over as few as 1000 pulses, is itself closer to the large-coherence limit -- again, compare the thick colored and thin black curves.

Figure~\ref{fig:Absorption}~(f) shows the evolution of the difference between the average response functions and reference at 10~eV with increasing coherence width. Unsurprisingly, we observe the largest difference at zero coherence width, followed by an overall smooth convergence as the coherence width is increased. Already at $\approx5$~eV coherence width, the average response function is only marginally different from the reference at 10~eV. Above~10~eV, we have observed that the average response function becomes virtually independent of the coherence width (not shown), which motivates our choice of using the 10~eV results as our reference. As we observed in panels (d,e), for each coherence width, the difference is largest around 50.31~eV (dotted line) and is associated with the gradual increase of the response function maximum with increasing coherence width. We attribute this to our observation from section~\ref{statistics}, that an increased coherence width leads to increasing the average peak intensity of VCM pulses, therefore strengthening the magnitude of the response function. 

\section{Conclusion}\label{conclusion}
In this paper, we introduced the variable coherence model (VCM) for phenomenologically  simulating SASE-FEL pulses with a continuously controllable coherence width, from minimally coherent as in the partial coherence model of Ref.~\cite{Pfeifer:10} up to fully coherent like in Fourier-limited pulses. Our approach also provides a straightforward avenue for including pulse shaping efforts~\cite{Li2024, Robles2025}. Specifically, we investigated the behavior of VCM pulses across three regimes of laser parameters akin to those available at the LCLS-I, LCLS-II, and FLASH FELs. We performed statistical analyses for the number and intensities of sub-pulses that comprise VCM pulses, both in time and frequency domains, and further investigated their correlation between the two domains. In both time and frequency, we found that the sub-pulse intensity distribution is maximally spread at zero coherence width and progressively narrows and converges towards the Fourier limit as the coherence width is increased. Similarly, we found that the sub-pulse number is maximized when the coherence width is zero and converges towards a single (Fourier limited) pulse for large coherence widths. Except for when the zero-coherence width average duration is already short compared to the available bandwidth, like for LCLS-II, we observed only marginal correlation between the time and frequency domains at small coherence widths for both the sub-pulse number and intensity.

We illustrated the effect of variable coherence width in nonlinear absorption for a multi-active electron atomic model. We showed that small coherence widths require a large number of VCM pulses for the average response function to converge, and that this number decreases with increasing coherence width. We also found that the average response function at zero or small coherence widths is notably different from that of fully coherent Fourier-limited pulse, highlighting the necessity of accounting for the stochastic properties of SASE-FEL pulses in nonlinear spectroscopy.


\begin{backmatter}
\bmsection{Acknowledgment}
The authors thank James Cryan for enlightening discussions. This material is based upon work supported by the National Science Foundation under Grant No.~PHY-2207656.

\bmsection{Disclosures} The authors declare no conflicts of interest.

\bmsection{Data Availability Statement} Data underlying the results presented in this paper are not publicly available at this time but may be obtained from the authors upon reasonable request.

\bmsection{Supplemental document} Additional sub-pulse statistics for the LCLS-I and LCLS-II sets of parameters of table~\ref{tab:FEL_table}.

\end{backmatter}


\newpage
\appendix

\renewcommand\thefigure{S\arabic{figure}}
\setcounter{figure}{0}    

\noindent\textbf{\Huge Supplemental document}

\section*{Additional sub-pulse statistics}\label{more_stats}
\begin{figure}[htbp]
    \centering\includegraphics[width=0.875\linewidth]{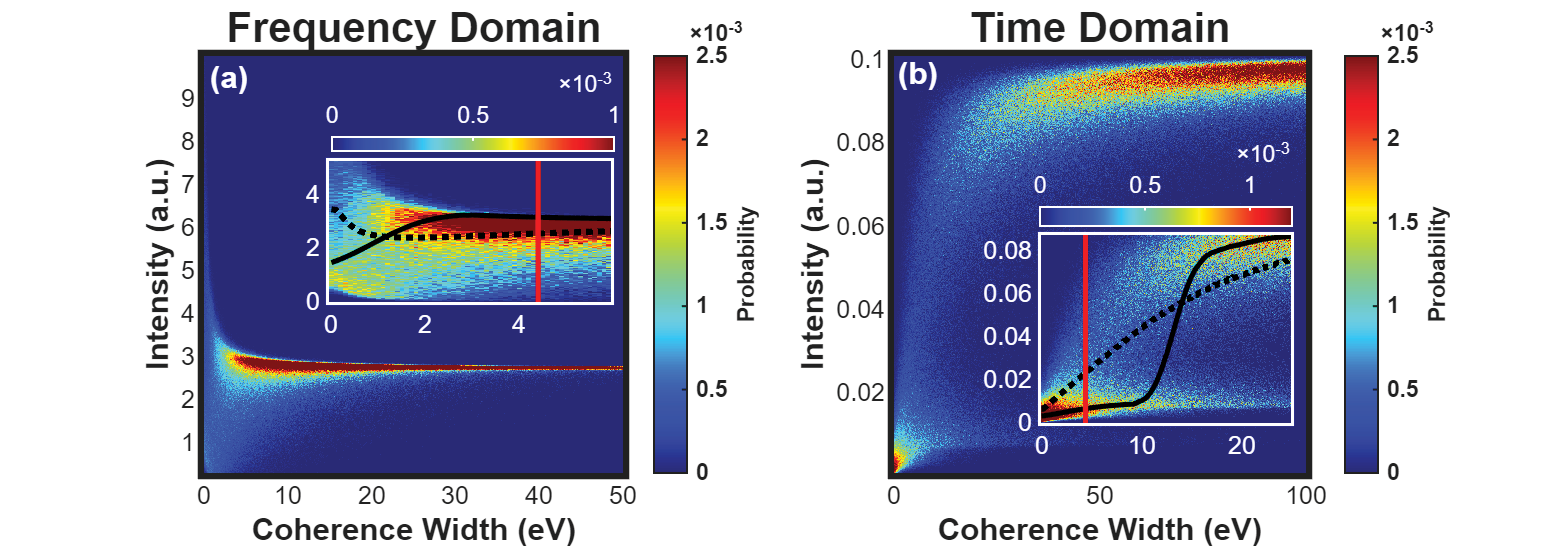}
    \caption{Same as figure 2 (main document) for the LCLS-I parameter set of table 1 (main document): simulated sub-pulse intensity distributions as a function of the coherence width in the (a) frequency and (b) time domains.}
    \label{fig:lcsl1_intdist}
\end{figure}

Figure~\ref{fig:lcsl1_intdist} shows the simulated sub-pulse intensity distributions as a function of the coherence width for the LCLS-I parameter set. As mentioned in section 3.1 of the main document, the LCLS-I case demonstrates similar behavior to the FLASH case within both domains. In the frequency domain, we see a large initial spread of possible sub-pulse intensities that begins to increase towards an asymptotic value as the coherence width is increased. In the time domain, the low-coherence width intensities concentrate towards lower values, but also increase towards an asymptotic value for large coherence widths. 

\begin{figure}[htbp]
    \centering\includegraphics[width=0.875\linewidth]{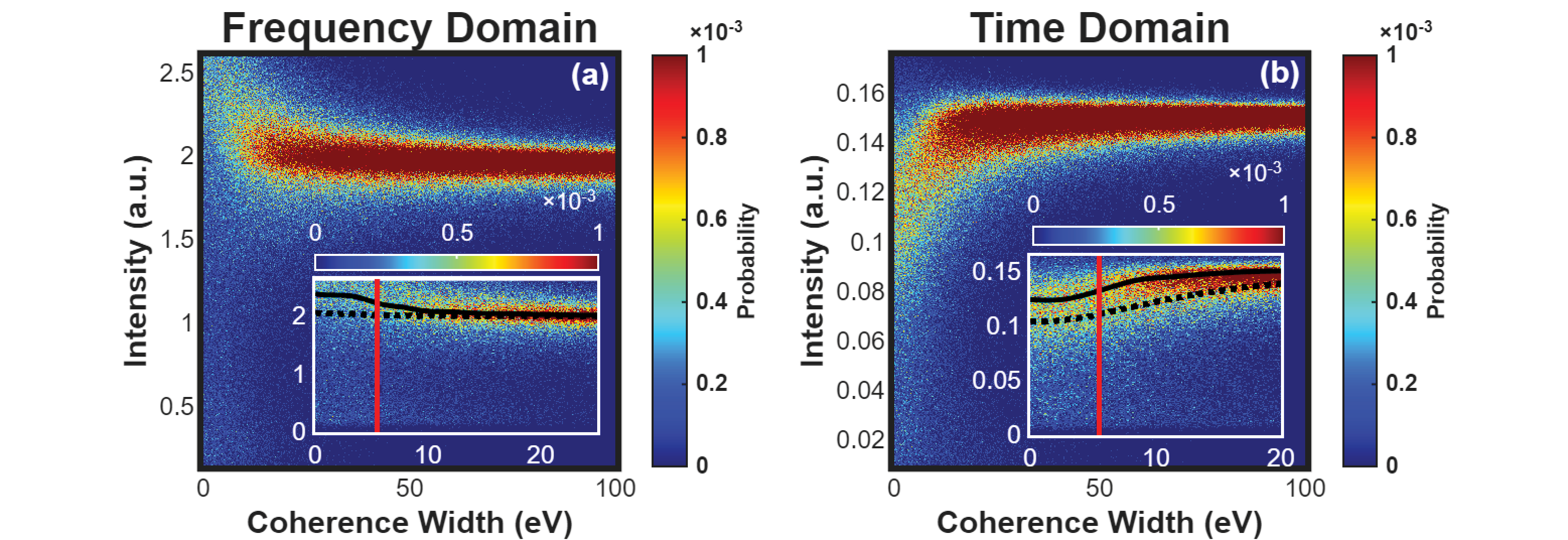}
    \caption{Same as figure 2 (main document) for the LCLS-II parameter set of table 1 (main document): simulated sub-pulse intensity distributions as a function of the coherence width in the (a) frequency and (b) time domains.}
    \label{fig:lcsl2_intdist}
\end{figure}

Contrary to both the FLASH and LCLS-I cases, the simulated sub-pulse intensity distributions for the LCLS-II case shown in figure~\ref{fig:lcsl2_intdist} have distinct behavior. In the frequency domain, we find that the most likely sub-pulse intensity is maximized at zero coherence width, and decreases towards an asymptotic value for larger coherence widths. In the time domain, we see an initial concentration towards lower intensities, and a gradual rise to an asymptotic intensity. This behavior in the time domain is reminiscent of the FLASH and LCLS-I cases, although the growth towards this asymptotic value for the LCLS-II case spans a smaller range of intensities in comparison.

\begin{figure}[htbp]
    \centering\includegraphics[width=0.875\columnwidth]{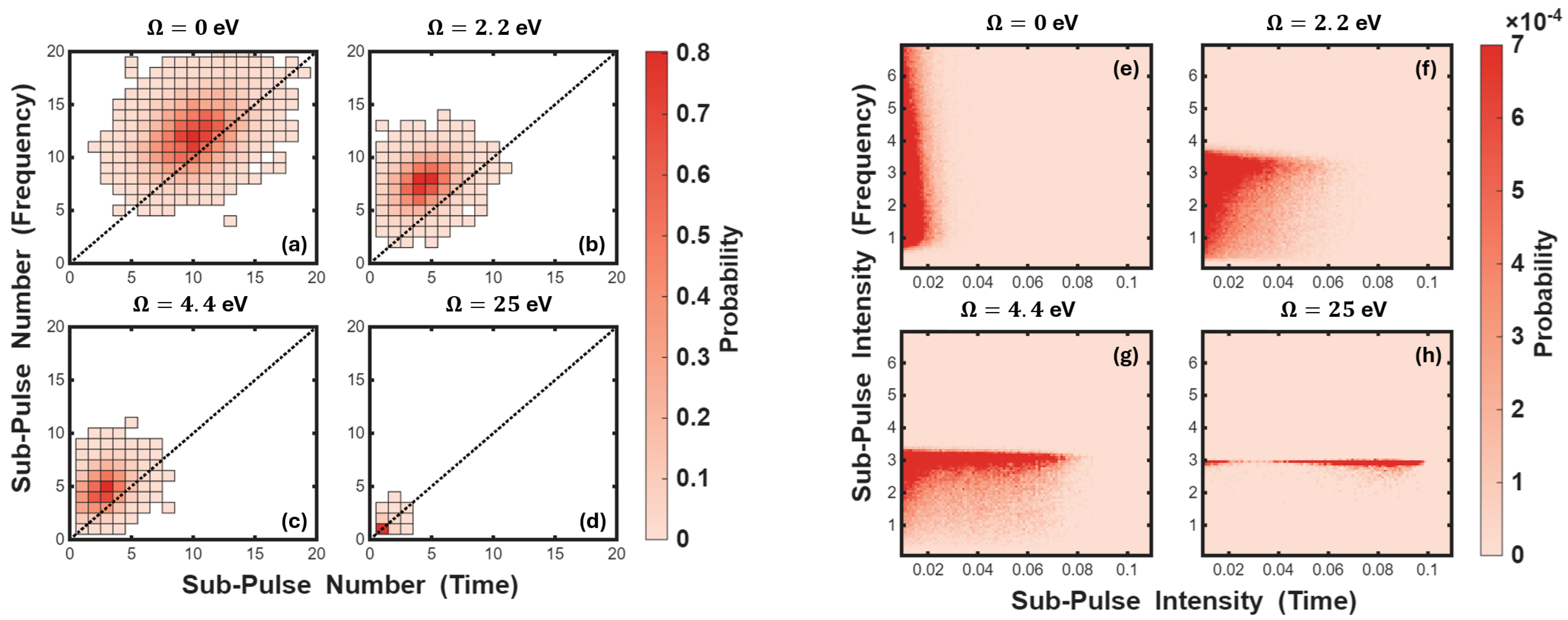} 
    \caption{Same as figure 4 (main document) for the LCLS-I parameter set of table 1 (main document): simulated joint probability distributions for the sub-pulse number (a-d) and intensity (e-h) in both the frequency and time domains.}
    \label{fig:JointHisto_lcls1}
\end{figure}

The simulated joint probability distributions discussed in section 3.2 of the main document are shown in figure~\ref{fig:JointHisto_lcls1} for the LCLS-I case. Again we obtain similar results to the FLASH parameter set for both the sub-pulse number and intensities. We see in figure~\ref{fig:JointHisto_lcls1} (a-d) that the joint probability distribution for the sub-pulse number demonstrates weak positive correlation for low coherence widths. Here as well, we attribute the bias of the distribution center towards the frequency domain to our choice of a higher prominence criterion in time in our sub-pulse detection -- see section 2.2 of the main document. As the coherence width increases, the strength of correlation increases along with it, while the overall number of sub-pulses decreases. In figure~\ref{fig:JointHisto_lcls1} (e-h), we find that the sub-pulse intensities between domains exhibit correlation in the same way that they did in the FLASH case, where we found that smaller intensities in time implied a greater likelihood of higher intensities in frequency, particularly for intermediate coherence widths.

\begin{figure}[htbp]
    \centering\includegraphics[width=0.875\linewidth]{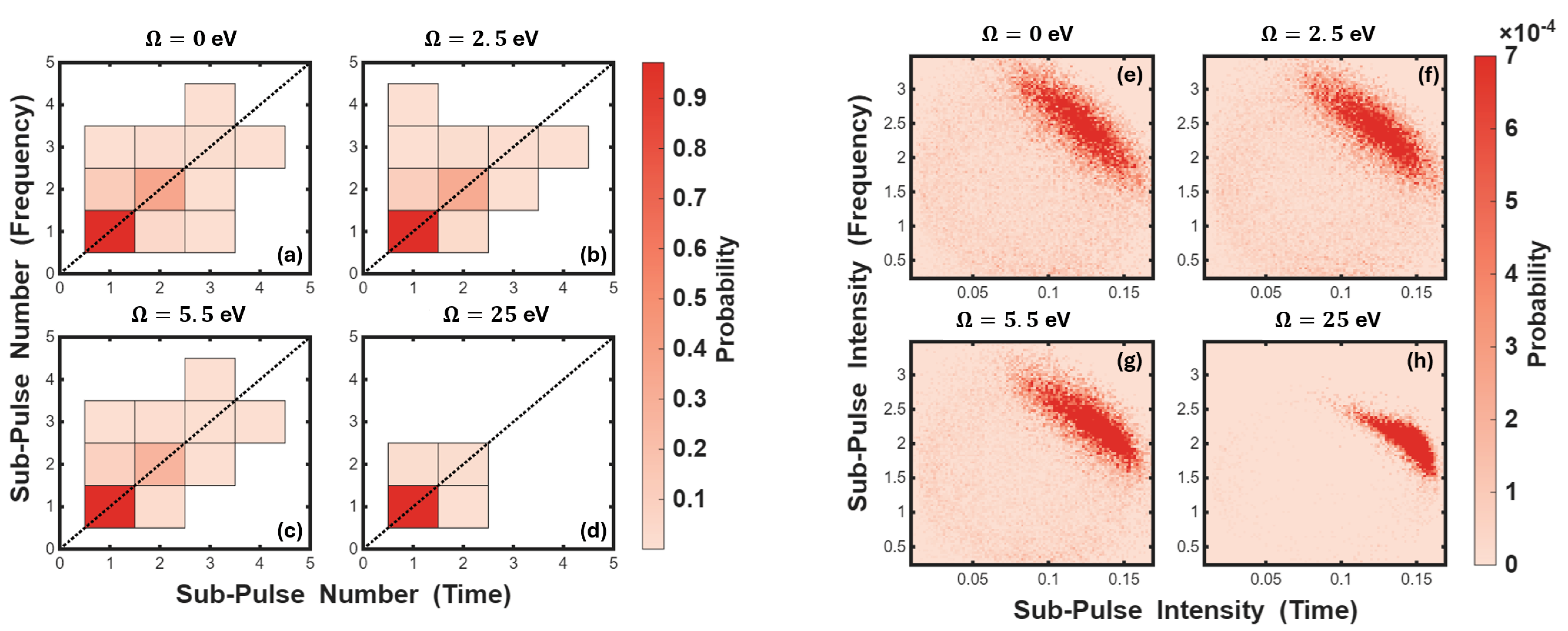}
    \caption{Same as figure 4 (main document) for the LCLS-II parameter set of table 1 (main document): simulated joint probability distributions for the sub-pulse number (a-d) and intensity (e-h) in both the frequency and time domains.}
    \label{fig:JointHisto_lcls2}
\end{figure}

The simulated joint probability distributions for the LCLS-II case are shown in figure~\ref{fig:JointHisto_lcls2}. The sub-pulse number distributions in figure~\ref{fig:JointHisto_lcls2} (a-d) exhibit considerable positive correlation for all coherence widths shown. For the sub-pulse intensity distributions shown in ~(e-h), we find that the coherence width has relatively little effect on the sub-pulse intensities between domains, as the sub-pulse intensities tend to cluster very close to their asymptotic values regardless of the coherence width, as demonstrated in figure~\ref{fig:lcsl2_intdist}. 

\end{document}